\documentstyle[prb,aps]{revtex}
\def\saff{{\it saff}}
\begin{document}

\title{Quantum Superintegrability and Exact Solvability in N Dimensions}
\author{Miguel A. Rodr\'{\i}guez\footnote{Electronic mail:
rodrigue@eucmos.sim.ucm.es}}
\address{Dept. de F\'{\i}sica Te\'orica II, Facultad
de F\'{\i}sicas, Universidad Complutense, 28040-Madrid, Spain}
\author{Pavel Winternitz\footnote{Electronic mail:
wintern@crm.umontreal.ca}}
\address{
Centre de Recherches Math\'ematiques and D\'epartement de Math\'ematiques
et de Statistique\\  Universit\'e de Montr\'eal, C.P. 6128, succ.
Centre-ville, Montr\'eal, Qu\'ebec H3C 3J7, Canada}
\date{\today}
\maketitle

\begin{abstract}A family of maximally superintegrable systems
containing the  Coulomb atom as a special case is constructed in
$n$-dimensional Euclidean  space. Two different sets of $n$ commuting
second order operators are found,  overlapping in the Hamiltonian alone.
The system is separable in several  coordinate systems and is shown to be
exactly solvable. It is solved  in terms of classical orthogonal
polynomials. The Hamiltonian and
$n$  further operators are shown to lie in the enveloping algebra of a
hidden  affine Lie algebra.
\end{abstract}

\section{Introduction}

We shall consider the stationary Schr\"odinger equation
\begin{equation}H\psi=E\psi,\quad H=-{1\over
2}\sum_{i=1}^{n}{\partial^2\over
\partial  x_i^2}+V(x_1,\ldots,x_n)\end{equation}
in an $n$-dimensional Euclidean space $E_n$. By analogy with classical 
Hamiltonian mechanics, such a system is called integrable if there 
exist $n-1$ algebraically independent linear
operators $X_a$ satisfying 
\begin{equation}
[H,X_a]=0,\quad [X_a,X_b]=0,\quad a,b=1,\ldots, n-1.
\end{equation} 
The system is called ``superintegrable" if there exist
$k$ further  operators $\{Y_1,\ldots, Y_k\}$ commuting with the
Hamiltonian
\begin{equation}
[H,Y_j],\quad j=1,\ldots k,
\end{equation}
such that the set $\{H,X_1,\ldots,X_n,Y_1,\ldots,Y_k\}$ is
algebraically  independent. Note that the additional operators $Y_i$
need not commute  with the operators $X_a$ nor amongst each other. The
number of additional  operators satisfies
\begin{equation}
1\le k\le n-1.
\end{equation}
For $k=1$ we call the system ``minimally superintegrable", for $k=n-1$
it  is ``maximally superintegrable".

The best known superintegrable systems in $E_3$ (and also in $E_n$ for 
any $n\ge 2$) are the harmonic oscillator and the hydrogen atom (or Kepler 
system in classical mechanics). The harmonic oscillator is superintegrable 
because of the $su(n)$ algebra of first and second order operators
commuting with  the Hamiltonian \cite{JH40,MS96}. The hydrogen atom in
$E_n$ is superintegrable,  because of the $o(n+1)$ Lie algebra of linear
operators commuting with the  Hamiltonian \cite{Pa34,Fo35,Ba36,En72,Mc72}.
In both cases it is possible to choose different subsets of 
$n$ operators commuting with each other and overlapping only in the 
Hamiltonian. Each subset corresponds to the separation of variables in the 
Schr\"odinger equation in a different system of coordinates \cite{KM76}.

Characteristic features of these two superintegrable systems are:
\begin{enumerate}
\item In classical mechanics all finite (bounded) trajectories are
periodic. Moreover, Bertrand's theorem \cite{Be73,Go90} tells us that
$\gamma/r$ and $\gamma\, r^2$ are the only spherically symmetric potentials
for which all finite trajectories are periodic.
\item In quantum mechanics these two systems are exactly solvable: their
energy levels can be calculated algebraically, as can the degeneracies of
these levels. Their eigenfunctions are polynomials in the appropriate
variables, multiplied by some overall factor.
\item These systems are extremely important in physical applications, both
in classical and quantum physics.
\end{enumerate}

It makes sense to search systematically for superintegrable systems in
classical and quantum mechanics, specially for maximally superintegrable
ones. It can be safely assumed that they will all have the above
properties 1 and 2 and hoped that they will also, to some degree, share
property 3.

In searches for superintegrable systems restrictions are imposed on the
form of the commuting operators $X_a$ and $Y_i$. A systematic search in
$E_2$ and $E_3$ was conducted some time ago
\cite{FM65,WS67,MS67,Ev90a,Ev91,Ev90b}. The restriction was that all
operators involved should be at most of second order. All superintegrable
systems satisfying this restriction in
$E_2$ and $E_3$ were found \cite{FM65,WS67,MS67,Ev90a,Ev91,Ev90b}. Four
classes of them exist in
$E_2$, 5 maximally superintegrable ($2n-1=5$ operators commuting with
$H$) and 8 minimally superintegrable ones ($n+1=4$ operators) in $E_3$.
These results have been recently extended to two and three dimensional
spaces of constant curvature and to complex spaces
\cite{KM96,KM00a,KM00b,KM01} and also to
certain two dimensional spaces of nonconstant
curvature \cite{KK01}.

With the restriction to second order operators all superintegrable
systems turned out to be multiseparable, that is, separable in at least
two different coordinate systems. In two dimensional spaces they also
turned out to be exactly solvable \cite{TT01}. By this we mean that their
energy spectra can be calculated algebraically (by solving algebraic
equations only) \cite{TT01,Tu94,Tu88}. It was also shown that
superintegrable systems are obtained by considering non-Abelian algebras
of generalized Lie symmetries \cite{ST01}.

The purpose of this article is to consider a family of integrable systems
in $n$ dimensional Euclidean space for any $n$. The family, containing
the $n$ dimensional hydrogen atom as a special case, is introduced in
Section 2, together with a set of $2n-1$ algebraically independent
operators, commuting with the Hamiltonian. In Section 3 we solve the
Schr\"odinger equation in parabolic
and spherical coordinates and show that it is
exactly solvable in a precise and well defined
sense
\cite{TT01,Tu94,Tu88}. Finally, in Section 4 we introduce parabolic
rotational coordinates in
$E_n$ and solve the Schr\"odinger equation in these coordinates
and also in spherical ones. We also prove the
exact solvability in this case. Some
conclusions are drawn in Section 5.


\section{A family of maximally superintegrable systems in $E_n$
containing the hydrogen atom}

Let us first consider the hydrogen atom in $n$ dimensional Euclidean
space $E_n$
\begin{equation}
H=-\frac{1}{2}\Delta-\frac{\gamma}{r},\quad
\Delta=\sum_{i=1}^n\frac{\partial^2}{\partial x_i^2},\quad
r=(x_1^2+\cdots +x_n^2)^{1/2}.\label{coul}
\end{equation}
This Hamiltonian commutes with $n(n+1)/2$ linear operators, namely
\begin{eqnarray}
L_{ik} & = & x_i\frac{\partial}{\partial x_k}-x_k\frac{\partial}{\partial
x_i},\quad 1\le i< k\le n\nonumber\\
A_i & = &\frac{1}{2}\sum_{a=1}^n(p_a L_{ia}+ L_{ia} p_a)+\gamma
\frac{x_i}{r}\quad 1\le i \le n
\end{eqnarray}
with $p_i=\partial_{x_i}$. The operators $L_{ik}$ correspond to angular
momentum, $A_i$ to the $n$ dimensional Laplace-Runge-Lenz vector,
characterizing the Coulomb or Kepler problem \cite{En72,Go90}. Only
$2n-1$ of the operators $\{H,L_{ik},A_i\}$ can be, and are, algebraically
independent. They satisfy the commutation relations
\begin{eqnarray}
&&[H,L_{ik}]=[H,A_i]=0\nonumber \\
&&[L_{ij},L_{ab}]=\delta_{ja} L_{ib}+\delta_{ib}L_{ja}-
\delta_{ia}L_{jb}-\delta_{jb} L_{ia}\nonumber\\
&&[L_{ij},A_k]=\delta_{jk}A_i-\delta_{ik} A_j\label{com}\\
&&[A_i,A_j]=-2 H L_{ij}\nonumber
\end{eqnarray}

The commutation relations (\ref{com}) in general correspond to a Kac-Moody
algebra \cite{DS93,DD98}. For a fixed energy, $H=E$ they correspond to
the Lie algebra of the rotation group $O(n+1)$, the Lorentz group $O(n,1)$
and the Euclidean group $E(n)$ for $E<0$, $E>0$ and $E=0$,
respectively. These symmetries for $n=3$ were discovered implicitly by
Pauli \cite{Pa34} and explicitly by Fock \cite{Fo35} and Bargmann
\cite{Ba36}.

According to the operator approach to the separation of variables
\cite{WF65,WL68,PW73,MP91,Mi97,Ka86}, separation of variables in
Schr\"odinger equation is achieved by looking for eigenfunctions of a
complete set of $n$ commuting second order operators
$\{H,X_i,\ldots,X_{n-1}\}$
\begin{equation}
H\psi=E\psi\quad X_a\psi=\lambda_a \psi,\quad
a=1,\ldots,n-1
\end{equation}
The operators $X_a$ will be at most linear in
$A_i$ and bilinear in
$L_{ik}$. If more than one inequivalent set of commuting operators
exists, the system is multiseparable, i.e., separable in more than one
coordinate system.

In view of the commutation relations (\ref{com}) any set of commuting
operators $\{X_i\}$ can contain at most one operator involving $A_i$:
\begin{equation}
X=\sum_{i}a_iA_i+\sum_{i,k,j,m}b_{ik,jm}L_{ik} L_{jm},\quad
\sum_{i=1}^n a_i^2\neq 0
\end{equation}
The complete sets of commuting operators can be classified under the action
of $O(n)$; in particular we can rotate and 
normalize so as to have $a_n = 1$, $a_k = 0$ for $k= 1,\ldots, n-1$. Here
we just give the example of the case $n=3$. It is easy to verify by a
direct calculation that in this case, precisely four inequivalent sets
exist:
$\{H,X_1,X_2\}$ with
\begin{eqnarray}
X_1=A_3,& &\quad  X_2=L_{12}^2\\
 X_1=A_3+a(L_{12}^2+L_{23}^2+L_{31}^2),& &\quad  X_2=L_{12}^2\\
 X_1=L_{12}^2+L_{23}^2+L_{31}^2,& &\quad X_2= L_{12}^2\\
 X_1=L_{12}^2+L_{23}^2+L_{31}^2,&  &\quad X_2=L_{23}^2+f L_{31}^2
\end{eqnarray}
They correspond to the separation of
variables in parabolic rotational coordinates, shifted spheroidal
coordinates, spherical coordinates and spheroconical coordinates,
respectively.

In each coordinate system it is possible to add further terms to the
potential $-\gamma/r$ in such a manner that the Schr\"odinger equation
still separates. The system will remain integrable and the corresponding
operators $X_1$ and $X_2$ will only be modified by the addition of a
scalar function, It is also possible to preserve superintegrability and to
require that the extended potentials should allow separation of variables
in at least two coordinate systems.

Here we will be interested in the most general potential allowing
separation of variables in the same four coordinate systems as the
hydrogen atom itself. In $E_3$ there is, up to equivalence, only one such
Hamiltonian, namely (see Ref.\cite{MS67,Ev90a})

\begin{equation}
H=-\frac{1}{2}\Delta
-\frac{\gamma}{r}+\frac{\beta_1}{x_1^2}+\frac{\beta_2}{x_2^2}\label{extend}
\end{equation}

One triplet of commuting operators for the Hamiltonian (\ref{extend})
consists of
\begin{eqnarray}
X & = &\frac{1}{2}(p_1L_{31}+L_{31}p_1+p_2L_{32}+L_{32}p_2)+2x_3\left(
\frac{\gamma}{2r}-\frac{\beta_1}{x_1^2}-\frac{\beta_2}{x_2^2}\right)
\nonumber\\ 
Z & = & L_{12}^2-2r^2\left(
\frac{\beta_1}{x_1^2}+\frac{\beta_2}{x_2^2}\right)\label{invthree}
\end{eqnarray}
Another triplet can be chosen to be $H$ and
\begin{eqnarray}
Y_1 & = &L_{12}^2+L_{23}^2+L_{31}^2-2r^2\left(
\frac{\beta_1}{x_1^2}+\frac{\beta_2}{x_2^2}\right)
\nonumber\\ 
Y_2 & = & L_{23}^2-2\beta_2
\frac{x_2^2+x_3^2}{x_2^2}\label{triplet}
\end{eqnarray}
It is the set of 5 algebraically independent operators
$\{H,X_1,X_2,Y_1,Y_2\}$ which guarantees that the Hamiltonian
(\ref{extend}) is maximally superintegrable.

The generalization to the $n$ dimensional Euclidean space $E_n$
is immediate. Thus, the Hamiltonian will be
\begin{equation}
H=-\frac{1}{2}\Delta
-\frac{\gamma}{r}+\sum_{i=1}^{n-1} \frac{\beta_i}{x_i^
2}\label{ndim}
\end{equation}
with $\Delta$ and $r$ as in Equation (\ref{coul}). One of the two
different complete sets of commuting operators can be chosen to be $H$ and
\begin{eqnarray}
X & = & \frac{1}{2}\sum_{k=1}^{n-1} (L_{nk} p_k+ p_k L_{nk} )+2x_n\left(
\frac{\gamma}{2r}-\sum_{i=1}^{n-1}\frac{\beta_i}{x_i^2}\right)\nonumber \\
Z_l & = & \sum_{1\le i <k \le l+1} L_{ik}^2 -2\left(\sum_{i=1}^{l+1}
x_i^2\right)
\left(\sum_{k=1}^{l+1} \frac{\beta_k}{x_k^2}\right),\quad 1\le l\le n-2
\label{comma}
\end{eqnarray}
Another complete set of commuting operators is again $H$ and
\begin{equation}
Y_p  =  \sum_{p\le i <k \le n} L_{ik}^2
-2\left(\sum_{i=p}^{n} x_i^2\right)
\left(\sum_{k=p}^{n-1} \frac{\beta_k}{x_k^2}\right),\quad 1\le p\le n-1
\label{commb}
\end{equation}
The two sets (\ref{comma}) and (\ref{commb}) are disjoint. If we set
$\beta_i=0,\; 1\le i\le n-1$, then the operator $Z_l$ will be a
Casimir operator of the group $O(l+1)$ acting on the coordinates
$\{x_1,\ldots, x_{l+1}\}$. The operator $Y_p$ will be a Casimir
operator of $O(n+1-p)$ acting on the coordinates $\{x_p,\ldots,x_n\}$

It is the Hamiltonian (\ref{ndim}) that we shall study in the following
sections, first for $n=3$, then for arbitrary $n$.

\section{Exact solvability of the superintegrable system for
$n=3$}

\subsection{Solution by separation of variables}

Let us first consider the Hamiltonian (\ref{extend}) and the complete set
of commuting operators (\ref{invthree}). We are looking for eigenvalues
and common eigenfunctions of the systems:
\begin{equation}
H\psi=E\psi,\quad X\psi=\lambda \psi,\quad Z\psi=k\psi\label{sepeq}
\end{equation}
To do this we introduce parabolic rotational coordinates, putting
\begin{equation}
x_1=\mu \nu \cos \phi,\quad x_2=\mu\nu\sin\phi,\quad
x_3=\frac{1}{2}(\mu^2-\nu^2)
\end{equation}
In these coordinates the operators in (\ref{sepeq}) are
\begin{eqnarray}
H &= & -\frac{1}{2(\mu^2+\nu^2)} \left(\frac{\partial^2}{\partial\mu^2}
+\frac{1}{\mu}\frac{\partial}{\partial\mu}
+\frac{\partial^2}{\partial\nu^2}+\frac{1}{\nu}\frac{\partial}{\partial
\nu}+4\gamma\right)-\nonumber\\ & &
\frac{1}{2\mu^2\nu^2}\left(\frac{\partial^2}{\partial\phi^2}- 
\frac{2\beta_1}{\cos^2\phi}-\frac{2\beta_2}{\sin^2\phi}\right)
\label{hampar}\\ X &= & \frac{1}{2(\mu^2+\nu^2)}
\left(-\nu^2\left( \frac{\partial^2}{\partial\mu^2}
+\frac{1}{\mu}\frac{\partial}{\partial\mu}\right)
+\mu^2\left(\frac{\partial^2}{\partial\nu^2}+
\frac{1}{\nu}\frac{\partial}{\partial
\nu}\right)+2\gamma(\mu^2-\nu^2)\right)+\nonumber\\ & &
\frac{\mu^2-\nu^2}{2\mu^2\nu^2}\left(\frac{\partial^2}{\partial\phi^2}- 
\frac{2\beta_1}{\cos^2\phi}-\frac{2\beta_2}{\sin^2\phi}\right)\nonumber\\
Z &= & \frac{\partial^2}{\partial\phi^2}- 
\frac{2\beta_1}{\cos^2\phi}-\frac{2\beta_2}{\sin^2\phi}\nonumber
\end{eqnarray}
We see immediately that the variables separate and we can solve the
corresponding ODE's to obtain
\begin{eqnarray}
\psi_{N_1,N_2,J} & =& (\sin\phi)^{p_2}(\cos\phi)^{p_1}  (\mu\nu)^m
e^{-\sqrt{-E/2}(\mu^2+\nu^2)}\times\nonumber \\ &&
P_J^{(p_2-1/2,p_1-1/2)}(\cos 2\phi)L_{N_1}^m (\sqrt{-2E}\mu^2)
L_{N_2}^m(\sqrt{-2E}\nu^2)\label{sol}
\end{eqnarray}
where $P^{(\alpha,\beta)}_J(z)$ and $L_N^m(x)$ are Jacobi and Laguerre
polynomials respectively. We have put
$$
\beta_i=\frac{1}{2}p_i(p_i-1),\quad m=2J+p_1+p_2 
$$
and the eigenvalues in Equation (\ref{sepeq}) are equal to
\begin{eqnarray}
E&=&-\frac{\gamma^2}{2(N_1+N_2+2J+p_1+p_2+1)^2}\nonumber\\
\lambda&=&-\frac{\gamma(N_1-N_2)}{N_1+N_2+2J+p_1+p_2+1}\\
k&=&-m^2=-(2J+p_1+p_2)^2\nonumber
\end{eqnarray}
We see that the bound state energy is given by a shifted Balmer
formula and the only effect of the $\beta_i/x_i^2$ terms in the potential
is to add a constant $p_1+p_2$ to the principal quantum number. The
solutions (\ref{sol}) are square integrable and correspond to bound
states when $J$, $N_1$ and $N_2$ are integers. They are polynomials
multiplied by a factor which, however, is not
``universal". It depends on the energy $E$ and
also on the angular quantum number $J$ (since
we have
$m=2J+p_1+p_2$).

The second set of commuting operators, namely (\ref{triplet}) also
corresponds to the separation of variables, this time in spherical
coordinates, chosen as
\begin{equation}
x_1=r\cos \theta,\quad x_2=r\sin\theta\cos\alpha,\quad
x_3=r\sin\theta\sin\alpha\label{spher}
\end{equation}
In these coordinates we have
\begin{eqnarray}
H\psi &= & -\frac{1}{2}\left[\frac{\partial^2}{\partial
r^2}+\frac{2}{r}\frac{\partial}{\partial
r}+\frac{1}{r^2}\left(\frac{\partial^2}{\partial \theta^2}+\cot \theta
\frac{\partial}{\partial\theta}+
\frac{1}{\sin^2\theta}\frac{\partial^2}{\partial
\alpha^2}\right)+\right.\nonumber\\ & &\left.\frac{2\gamma}{r}-
\frac{2}{r^2}\left(\frac{\beta_1}{\cos^2\theta}+\frac
{\beta_2}{\sin^2\theta\cos^2\alpha}\right)\right]\psi=E\psi \nonumber,\\ 
Y_1
\psi & = &
\left(\frac{\partial^2}{\partial \theta^2}+\cot \theta
\frac{\partial}{\partial\theta}-\frac{2\beta_1}{\cos^2\theta}+
\frac{k_2}{\sin^2\theta}\right)\psi=k_1\psi,\label{system}\\
Y_2 \psi & = & \left(\frac{\partial^2}{\partial \alpha^2}
-\frac{2\beta_2}{\cos^2\alpha}\right)\psi=k_2\psi.\nonumber
\end{eqnarray}

The coordinates separate and we obtain $\psi=R(r)F(\theta)G(\alpha)$
where $F$ and $G$ can be expressed in terms of Jacobi polynomials and $R$
in terms of Laguerre ones.

The explicit expression for the eigenfunctions in this type of spherical
coordinates is:
\begin{eqnarray}\psi_{N,J_1,J_2}(r,\theta,\alpha)&= &
r^{m_1-1/2}e^{-\sqrt{-2E}\, r} L_{N}^{2m_1}(2\sqrt{-2E}\, r)
(\sin\theta)^{m_1}(\cos\theta)^{p_1}(\cos\alpha)^{p_2}\times \nonumber\\ &&
P_{J_1}^{(m_2,p_1-1/2)}(\cos 2\theta)  P^{(-1/2,p_2-1/2)}_{J_2}(\cos
2\alpha)\end{eqnarray}
 with eigenvalues equal to:
\begin{eqnarray}
E & = & -\frac{\gamma^2}{2(N+2J_1+2J_2+p_1+p_2+1)^2}
\\k_1 & = & {1\over 4}-m_1^2,\quad k_2  =  -m_2^2,\quad
m_1=2J_1+2J_2+p_1+p_2+\frac{1}{2},\quad
m_2=2J_2+p_2\nonumber\end{eqnarray}

\subsection{Exact solvability and underlying affine Lie algebra}

We have established that the potential (\ref{extend}) provides a
Hamiltonian that is maximally superintegrable and multiseparable. Let us
now turn to the question of exact solvability. A Hamiltonian is exactly
solvable if its spectrum can be calculated algebraically. This occurs if
it can be explicitly transformed into block diagonal form where each block
is finite dimensional. This in turn means that there exists an infinite
flag of finite dimensional subspaces in the Hilbert space ${\cal H}$ of
bound state solutions that is preserved by the Hamiltonian
\cite{TT01,Tu94,Tu88}
\begin{equation}
{\cal H}_1\subset {\cal H}_2\subset\cdots\subset {\cal H},\quad H{\cal
H}_i\subseteq {\cal H}_i.
\end{equation}
Typically this occurs under the following circumstances
\begin{enumerate}
\item The bound state wave functions are polynomials in some variables,
possible multiplied by  some factor $g$, i.e.,
$\psi=g P$. We then have
\begin{equation}
h=g^{-1}H g,\quad hP=EP,
\end{equation}
that is, there exists a gauge transformation and a change of variables to
a new Hamiltonian
$h$ that has polynomial eigenfunctions.
\item The gauge transformed Hamiltonian $h$ is an element of the
enveloping algebra of some affine Lie algebra $L$, a basis of which can
be realized by the operators
\begin{equation}
K_i=\frac{\partial}{\partial s_i},\quad M_{ik}=s_i\frac{\partial}{\partial
s_k},\quad i,k\in\{1,\ldots, n\}
\end{equation}
(in some coordinates $s_i$).
\end{enumerate}

Let us now investigate the Hamiltonian (\ref{extend}) and the commuting set
of operators (\ref{invthree}) and (\ref{triplet}) from this point of view.

Consider the wave function (\ref{sol}) in parabolic coordinates. They do
have the required form
\begin{eqnarray}
\psi_{N_1,N_2,J} & = & gP_{N_1,N_2,J}(s,t,z), \quad s=\sqrt{-2E}\mu^2,\quad
t=\sqrt{-2E} \nu^2, \quad 
z  =\cos 2\phi\\ g& = &
\left(\frac{1+z}{2}\right)^{p_1/2}\left(\frac{1-z}{2}\right)^{p_2/2}
\left(\frac{s\,t}{-2E}\right)^{m/2} e^{-(s+t)/2}
\nonumber
\end{eqnarray}
where $P$ is a polynomial (a product of three polynomials
in one variable each).
To proceed further we must get rid of the conformal factor
$(\mu^2+\nu^2)^{-1}$ figuring in equation (\ref{hampar}) and replace the
system (\ref{sepeq}) by
\begin{equation}
Q_0\psi  =  2\gamma\psi,\quad  Q_1\psi=2\lambda\psi,\quad Z \psi  =
-m^{2}\psi\label{newsepeq}
\end{equation}
\begin{eqnarray*}
Q_0 & = & (\mu^2+\nu^2)(H-E) +2\gamma =\\ &&-\frac{1}{2}
\left(\frac{\partial^2}{\partial \mu^2}+\frac{1}{\mu}
\frac{\partial}{\partial
\mu} +\frac{\partial^2}{\partial \nu^2}
+\frac{1}{\nu}\frac{\partial}{\partial
\nu}\right)+\frac{m^2}{2}\left(\frac{1}{\mu^2}
+\frac{1}{\nu^2}\right)-E(\mu^2+\nu^2),\\ Q_1 & = & 2X+(\mu^2-\nu^2)(H-E)
=\\ && -\frac{1}{2}\left(\frac{\partial^2}{\partial
\mu^2}+\frac{1}{\mu}\frac{\partial}{\partial\mu}-
\frac{\partial^2}{\partial
\nu^2}-\frac{1}{\nu}\frac{\partial}{\partial\nu}\right)+
\frac{m^2}{2}\left(\frac{1}{\mu^2}-\frac{1}{\nu^2}\right)-
E(\mu^2-\nu^2),\\
Q_2 & = & Z.
\end{eqnarray*}
We see here a phenomenon which has been called ``metamorphosis"
\cite{HG84} or ``migration" \cite{TT01} of the coupling constant. In
equation (\ref{newsepeq}) the energy
$E$ plays  the role of the frequency of a harmonic oscillator whereas the
Coulomb coupling constant $\gamma$ plays the role of an eigenvalue of
$Q_0$. The other eigenvalues, $\lambda$ and $m^2$, remain eigenvalues (of
$Q_1$ and $Z$ respectively).

Similarly as in the case of potentials containing the Coulomb atom as a
special case in 2 dimensions, it is the system (\ref{newsepeq}) (rather
than the original system (\ref{sepeq})) that is exactly solvable in the
sense defined above. Indeed, let us gauge rotate the
operators $Q_0$, $Q_1$ and
$Q_2$ and transform to the variables $s$, $t$ and $z$. 
We obtain
\begin{eqnarray}
\tilde{Q}_0 & =& g^{-1}Q_0 g=-2\sqrt{-2E}
\left(s\partial_s^2+(m+1-s)\partial_s +t\partial_t^2+(m+1-t)\partial_t
-m-1\right),\nonumber\\
\tilde{Q}_1 & = & g^{-1} Q_1 g=
-2\sqrt{-2E}\left(s\partial_s^2+(m+1-s)
\partial_s-t\partial_t^2-(m+1-t)\partial_t\right),
\\
\tilde{Q}_2 & = & g^{-1}Q_2
g=4(1-z^2)\partial{z^2}+4(p_1-p_2-(p_1+p_2+1)z)\partial_z-(p_1+p_2)^2.
\nonumber
\end{eqnarray}
We see that $\tilde{Q}_{\mu},\; \mu=0,1,2$ lie in the enveloping algebra
of the direct sum of three affine Lie algebras,
$\saff(1,{\bf R})\oplus
\saff(1,{\bf R})\oplus \saff(1,{\bf R})$,
realized by
\begin{equation}
\{\partial_s,s\partial_s,\partial_t,t\partial_t,\partial_z,z\partial_z\}.
\label{alg}
\end{equation}
Now let us consider the two remaining operators $Y_1$ and $Y_2$ of
(\ref{system}). They are not diagonal in the basis that we use (where
$H,X,Z$ and equivalently $Q_0,Q_1,Q_2$ are diagonal). They do however
commute with the Hamiltonian so they can only mix states of equal energy.
We also have
\begin{equation}
[Y_1,Z]=0,\quad [Y_2,Z]\neq 0,\quad [Y_1,X]\neq 0.
\end{equation}

It follows that $Y_1$ also preserves the quantum number $k=-m^2$ of
equation (\ref{sepeq}), whereas $Y_2$ mixes all states of a given energy.
The gauge factor $g$ depends not only on the energy, but also on $m$, the
eigenvalue of $Z$. Thus, $\tilde{Y}_1=g^{-1} Y_1 g$ should transform
polynomials into polynomials, whereas $\tilde{Y}_2=g^{-1} Y_2 g$ is not
obliged to. Performing the gauge transformation and change of variables
we find
\begin{equation}
\tilde{Y}_1= g^{-1} Y_1 g =s\,t(\partial_s-\partial_t)^2
-(m+1)(s-t)(\partial_s -\partial_t) -m(m+1). \label{gauged}
\end{equation}

The algebra underlying this expression includes $t\partial_s$ and
$s\partial_t$, in addition to the elements listed in (\ref{alg}). We
recognize this to be the Lie algebra $\saff
(2,{\bf R})\oplus
\saff(1,{\bf R})$

Superintegrable systems, including the hydrogen atom as special case, in 2
dimensions were associated with the algebra $\saff(2,{\bf R})$. The
extension to
$n=3$ is seen to lead to $\saff(2,{\bf R})\oplus \saff(1,{\bf R})$, not to
$\saff(3,{\bf R})$ as one might have expected.
It follows from expression (\ref{gauged}) that $\tilde{Y}_1$ will take
polynomials into polynomials and indeed we have
\begin{eqnarray}
\tilde{Y}_1 P_{N_1,N_2,J} &= &-[2N_1N_2
+(N_1+N_2+m)(m+1)]P_{N_1,N_2,J}+\nonumber\\ &&(N_1+m)(N_2+1)
P_{N_1-1,N_2+1,J}+(N_1+1)(N_2+m)P_{N_1+1,N_2-1,J}.
\end{eqnarray}
The operator $\tilde{Y}_2$ does not take polynomials into polynomials and
cannot be written as an element of the enveloping algebra of an
affine Lie algebra.

The one-dimensional equations appearing above are easily shown to be
related to the standard types in the classification of exact and
quasi-exact solvable one-dimensional systems\cite{GK93,Sh89}.

\section{Exact solvability of the superintegrable system in
$E_n$}

\subsection{Solution by separation of variables}
We now consider the Hamiltonian (\ref{ndim}) for arbitrary $n$. It allows
separation of variables in many coordinate systems. We shall use
parabolic rotational coordinates corresponding to the set of operators
(\ref{comma}) and spherical ones corresponding to the set (\ref{commb}).
The parabolic coordinates $(\mu,\nu,\theta_1,\ldots\theta_{n-2})$ are
defined by the relations
\begin{eqnarray}
x_1 & = & \mu\nu\cos \theta_1\cos \theta_2\ldots
\cos\theta_{n-3}\cos\theta_{n-2}, \nonumber\\
x_2 & = & \mu\nu\cos \theta_1\cos \theta_2\ldots
\cos\theta_{n-3}\sin\theta_{n-2}, \nonumber\\
x_3 & = & \mu\nu\cos \theta_1\cos \theta_2\ldots
\sin\theta_{n-3}, \nonumber\\
&\vdots& \label{coord}\\
x_{n-2} & = & \mu\nu\cos\theta_1\sin\theta_2,\nonumber\\
x_{n-1} & = & \mu\nu\sin\theta_1,\nonumber\\
x_n & = & \frac{1}{2}(\mu^2-\nu^2).\nonumber
\end{eqnarray}

We put $\beta_i=p_i(p_i-1)/2$ in equation (\ref{ndim}). The eigenvalue
problem that we have to solve is
\begin{equation}
H\psi=E\psi,\quad X\psi=\lambda\psi,\quad Z_l\psi=k_l \psi,\quad 1\le l\le
n-2.
\end{equation}
In parabolic coordinates the operators $H$ and $X$ will involve all
variables, the operators $Z_l$ will involve the angles only.
Indeed, we have
\begin{eqnarray}
H & = & -\frac{1}{2(\mu^2+\nu^2)}\left(
\frac{\partial^2}{\partial
\mu^2}+\frac{n-2}{\mu}\frac{\partial}{\partial \mu}+
\frac{\partial^2}{\partial
\nu^2}+\frac{n-2}{\nu}\frac{\partial}{\partial
\nu}\right)
-\frac{2\gamma}{\mu^2+\nu^2}-  \frac{1}{2\mu^2\nu^2}\bigg[
\Delta(S_{n-2})-\nonumber\\
&& \frac{p_1(p_1-1)}{\cos^2\theta_1\cdots
\cos^2\theta_{n-2}}-\frac{p_2(p_2-1)}{\cos^2\theta_1\cdots
\cos^2\theta_{n-3}\sin^2\theta_{n-2}}-\cdots -
\frac{p_{n-1}(p_{n-1}-1)}{\sin^2\theta_1}\bigg],\label{hamn}\\ \nonumber\\
X & = & \frac{1}{2(\mu^2+\nu^2)}\left[
-\nu^2\left(\frac{\partial^2}{\partial
\mu^2}+\frac{n-2}{\mu}\frac{\partial}{\partial \mu}\right)+\mu^2\left(
\frac{\partial^2}{\partial
\nu^2}+\frac{n-2}{\nu}\frac{\partial}{\partial
\nu}\right)\right]
\nonumber\\ && +\gamma\frac{\mu^2-\nu^2}{\mu^2+\nu^2}+ 
\frac{\mu^2-\nu^2}{2\mu^2\nu^2}\bigg[
\Delta(S_{n-2})-\nonumber\\ && \frac{p_1(p_1-1)}{\cos^2\theta_1\cdots
\cos^2\theta_{n-2}}-\frac{p_2(p_2-1)}{\cos^2\theta_1\cdots
\cos^2\theta_{n-3}\sin^2\theta_{n-2}}-\cdots -
\frac{p_{n-1}(p_{n-1}-1)}{\sin^2\theta_1}\bigg]\label{xndim}
\end{eqnarray}
where $\Delta(S_{n-2})$ is
the Laplace operator on an $n-2$ dimensional sphere

The operators $Z_l$ satisfy
\begin{eqnarray*}
Z_1\psi=\left(\frac{\partial^2}{\partial\theta_{n-2}^2}-
\frac{p_1(p_1-1)}{\cos^2\theta_{n-2}}-
\frac{p_2(p_2-1)}{\sin^2\theta_{n-2}}\right)\psi=k_1\psi,\\
Z_2\psi=\left(\frac{\partial^2}{\partial\theta_{n-3}^2}-
\tan
\theta_{n-3}\frac{\partial}{\partial\theta_{n-3}}+
\frac{k_1}{\cos^2\theta_{n-3}}-
\frac{p_3(p_3-1)}{\sin^2\theta_{n-3}}\right)\psi=k_2\psi,
\end{eqnarray*}
and in general
\begin{eqnarray}
Z_l\psi &=&\left(\frac{\partial^2}{\partial\theta_{n-l-1}^2}-
(l-1)\tan\theta_{n-l-1}\frac{\partial}{\partial\theta_{n-l-1}}+
\frac{k_{l-1}}{\cos^2\theta_{n-l-1}}-
\frac{p_{l+1}(p_{l+1}-1)}{\sin^2\theta_{n-l-1}}\right)\psi=k_l\psi
\label{gen}\\
&& 1\le l\le n-2,\quad k_0=-p_1(p_1-1).\nonumber 
\end{eqnarray}

We write
\begin{equation}
\psi=M(\mu)N(\nu) \prod_{l=1}^{n-2} F_l(\theta_{n-l-1})
\end{equation}
and solve (\ref{gen}) to obtain $F_l$ in terms of Jacobi polynomials
\begin{equation}
F_l(\theta_{n-l-1})=(\sin\theta_{n-l-1})^{p_{l+1}}
(\cos\theta_{n-l-1})^{m_{l-1}+1-l/2}P_{J_l}^{(p_{l+1}-1/2,m_{l-1})} (\cos
2\theta_{n-l-1})
\end{equation}
\begin{equation}
m_l=2\sum_{i=1}^l J_i+\sum_{i=1}^{l+1} p_i +\frac{l-1}{2},\quad
k_l=\frac{(l-1)^2}{4}-m_l^2
\end{equation}
The equations for $M(\mu)$ and $N(\nu)$ are
obtained from equation (\ref{hamn}) and (\ref{xndim}) once the angular part
is replaced by $k_{n-2}$. The final result is that the wave functions are:
\begin{eqnarray}
\psi_{N_1,N_2,J_1,J_2,\ldots,J_{n-2}}(\mu,\nu,\theta_1,
\ldots,\theta_{n-2})
=(\mu\nu)^{\sigma} e^{-\sqrt{-E/2}(\mu^2+\nu^2)}\times \nonumber\\ 
\prod_{l=1
}^{n-2}(\sin\theta_{n-l-1})^{p_{l+1}}
(\cos\theta_{n-l-1})^{m_{l-1}+1-l/2}
L_{N_1}^{m_{n-2}}(\sqrt{-2E}\mu^2)L_{N_2}^{m_{n-2}}(\sqrt{-2E}\nu^2)
\times \nonumber \\ \prod_{l=1}^{n-2} P_{J_l}^{(p_{l+1}-1/2,m_{l-1})}
(\cos2\theta_{n-l-1}),\quad   \sigma=2\sum_{i=1}^{n-2}J_i+\sum_{i=1}^{n-1}
p_i 
\end{eqnarray}

The energy is given by a shifted Balmer formula
\begin{equation}
E=-\frac{\gamma^2}{2(N_1+N_2+2\sum_{i=1}^{n-2}J_i+\sum_{i=1}^{n-1}
p_i+\frac{n-1}{2})^2}
\end{equation}
and the remaining quantum number is
\begin{equation}
\lambda=-\frac{\gamma(N_1-N_2)}{N_1+N_2+2\sum_{i=1}^{n-2}J_i+
\sum_{i=1}^{n-1} p_i+\frac{n-1}{2}}
\end{equation}

We see that the case of $n$ arbitrary is a straightforward
generalization of $n=3$ and involves the same functions, namely, Jacobi
and Laguerre polynomials.

Obviously, one can also solve in spherical coordinates. In fact, 
formulas (\ref{coord}) can be written as
\begin{equation}
x_a=\mu\nu s_a,\quad x_n=\frac{1}{2}(\mu^2-\nu^2)\quad a=1,\ldots, n-1,
\quad \sum_{a=1}^{n-1} s_a^2 =1
\end{equation}
and we could introduce any coordinates on the $S_{n-2}$ sphere
that allow separation of variables in the Laplace-Beltrami equation. For a
discussion of such coordinate systems see \cite{Ka86,Vi68,VK65,IP99,PW01}.

We will write for the sake of completeness the
explicit expression of the eigenfunctions in
the following set of spherical coordinates on
the
$S_{n-1}$ sphere (which are a generalization
to dimension $n$ of those we used in the case
$n=3$, see Equation (\ref{spher})):
\begin{eqnarray}
x_1&= &r\cos\theta_1\nonumber\\ 
x_1&= &r\sin\theta_1\cos\theta_2\nonumber\\ 
&\vdots&\\  
x_{n-1}&= &r\sin\theta_1\cdots\sin\theta_{n-2}\cos\theta_{n-1}\nonumber\\ 
x_n&= &r\sin\theta_1\cdots\sin\theta_{n-2}\sin\theta_{n-1} \nonumber
\end{eqnarray}
and the Hamiltonian can be written in these coordinates as
\begin{eqnarray}H&=& 
-{1\over 2}\left[\partial_{r}^2+{n-1\over  r}\partial_r+{2\gamma\over
r}\right]- {1\over 2r^2}\bigg\{\bigg[ \partial^2_{\theta_1}+(n-2)\cot
\theta_1
\partial_{\theta_1}-{p_1(p_1 -1)\over \cos^2\theta_1}+ \nonumber \\ &&
{1\over
\sin^2\theta_1}\bigg[
\partial_{\theta_2}^2+(n-3)\cot\theta_2\partial_{\theta_2}-{p_2(p_2
-1)\over
\cos^2\theta_2}+ \nonumber \\ && \cdots +
{1\over\sin^2\theta_{n-3}}\bigg[\partial^2_{\theta_{n-2}}+\cot\theta_{n-2}
\partial_{\theta_{n-2}} -{p_{n-2}(p_{n-2} -1)\over \cos^2\theta_{n-2}}+
\nonumber\\ & & {1\over\sin^2\theta_{n-2}}\bigg[\partial^2_{\theta_{n-1}}
-{p_{n-1}(p_{n-1} -1)\over \cos^2\theta_{n-1}}\bigg]\cdots\bigg]\bigg\}
\end{eqnarray}

The set of $Y_l,\; l=1,\ldots, n-1$ operators are:
\begin{eqnarray}
Y_l&=&\partial^2_{\theta_l}+(n-l-1)\cot\theta_l\partial_{\theta_l}
-{p_{l}(p_{l} -1)\over \cos^2\theta_l}+{k_{l+1}\over
\sin^2\theta_l},\quad l=1,\ldots n-2\nonumber\\
Y_{n-1}&=&\partial^2_{\theta_{n-1}}
-{p_{n-1}(p_{n-1} -1)\over \cos^2\theta_{n-1}}\end{eqnarray}
and the eigenvalue equations:
\begin{equation}
H\psi=E\psi,\quad Y_lG_l(\theta_{l})=k_lG_l(\theta_{l}),\quad
\l=1,\ldots, n-1,\quad
\psi=R(r)\prod_{l=1}^{n-1}G_l(\theta_{l})
\end{equation}
can be easily solved. The solution for
the angular part is ($m_n=-1/2$):
\begin{equation}\prod_{l=1}^{n-1}G_l(\theta_{l})=\prod_{l=1}^{n-1}(\sin
\theta_{l})^{m_{l+1}+1-(n-l)/2}(\cos\theta_{l})^{p_{l}} 
P_{J_l}^{(m_{l+1},p_{l}-1/2)}(\cos 2\theta_{l})\end{equation}
and for the radial part:
\begin{equation}
R(r)=r^{m_{1}-(n-2)/2}
e^{-\sqrt{-2E}r}L_{N_r}^{2m_{1}}(2\sqrt{-2E}r)
\end{equation}
The energy is written as:
\begin{equation}E=-{\gamma^2\over 2(N_r+2\sum_{i=1}^{n-1}
J_i+\sum_{i=1}^{n-1}p_{n-i}+{1\over 2}(n-1))^2}
\end{equation}
and the eigenvalues of the operators $Y_l$ are:
\begin{equation}
k_l={1\over 4}(n-l-1)^2-m_l^2, \quad
m_l=2\sum_{i=l}^{n-1}J_{i}+\sum_{i=l}^{n-1} p_{i}+{1\over 2}(n-l-1),\quad
l=1,\ldots, n-1
\end{equation}
Finally, the eigenfunctions are:
\begin{eqnarray}
\psi_{N,J_1,\ldots,J_{n-1}}(r,\theta_1,\ldots,\theta_{n-1}) & = &
r^{m_{1}-(n-2)/2}
e^{-\sqrt{-2E}r}L_{N_r}^{2m_{1}}(2\sqrt{-2E}r)\times
\nonumber \\ & &
\prod_{l=1}^{n-1}\bigg[
(\sin\theta_{l})^{m_{l+1}+1-(n-l)/2}(\cos\theta_{l})^{p_{l}}
\nonumber \\ & & P_{J_l}^{(m_{l+1},p_{l}-1/2)}(\cos 2\theta_{l})\bigg]
\end{eqnarray}

\subsection{Exact solvability}
The exact solvability of the system (\ref{ndim}) for general $n$ can be
treated in the same way as for $n=3$. We can gauge transform each of the
operators in the set (\ref{comma})  separately and transform to the
variables
\begin{equation}
s=\sqrt{-2E} \mu^2,\quad t=\sqrt{-2E} \nu^2,\quad z_{n-l+1} =\cos
2\theta_{n-l+1},\quad l=1,\ldots, n-2
\end{equation}
Before doing this, we again introduce $Q_0$ and $Q_1$ as in
equation (\ref{newsepeq}).

The final result is
\begin{eqnarray}
\tilde{Q}_0+\tilde{Q}_1 &=&g^{-1}(Q_0+Q_1)g=
-2\sqrt{-2E}(2s\partial_s^2+2(1+m_{n-2}-s)\partial_s-m_{n-2}-1)
\nonumber \\
\tilde{Q}_0-\tilde{Q}_1&=&g^{-1}(Q_0-Q_1) g=-2\sqrt{-2E}
(2t\partial_t^2+2(1+m_{n-2}-t)\partial_t-m_{n-2}-1)\\
\tilde{Z}_l & = & g^{-1}Z_l g =4(1-z_{n-l-1}^2) 
\frac{\partial^2}{\partial
z_{n-l-1}^2}+\nonumber\\ &&
4\left(m_{l-1}-p_{l+1}+\frac{1}{2}-\left(p_{l+1}+m_{l-1}+\frac{3}{2}
\right)z_{n-l-1}\right)\partial_{z_{n-l-1}}+\nonumber\\
&&\frac{l(l-2)}{4}-(m_{l-1}+p_{l+1})(m_{l-1}+p_{l+1}+1)\nonumber
\end{eqnarray}

We see that the entire set of operators
$\{Q_0,Q_1,Z_1,\ldots,Z_{n-2}\}$ lies in the enveloping algebra
of direct product of $n$ special affine Lie algebras
$\saff(1,{\bf R})$.

Finally let us turn to the other complete set of commuting operators
(\ref{commb}), associated with the separation of variables in spherical
coordinates. Among these operators there is just one, namely $Y_1$, that
commutes with all the operators $Z_l$. We have
\begin{equation}
[Y_1,Z_l]=0,\quad [Y_1,X]\neq 0, \quad [Y_p,Z_l]\neq
0,\quad 1\le l\le n-2,\quad 2\le p\le n-1
\end{equation}
Thus, $\tilde{Y}_1$ will take polynomials into polynomials but
$\{\tilde{Y}_2,\ldots,\tilde{Y}_{n-1}\}$ will not. We have
\begin{eqnarray}
\tilde{Y}_1 & = & g^{-1}\tilde{Y}_1 g
=s\,t(\partial_s-\partial_t)^2-(m_{n-2}+1)(s-t)(\partial_s-\partial_t)-
\nonumber\\ && m_{n-2}(m_{n-2}+1)+\frac{(n-3)(n-1)}{4}
\end{eqnarray}

Finally we see that the ``hidden Lie algebra" that is not a symmetry
algebra of the problem, but underlies its exact solvability is
$\saff(2,{\bf R})\oplus [\saff(1,{\bf R})]_1 \oplus\cdots \oplus
[\saff(1,{\bf R})]_{n-2}$ generated
 by 
\begin{equation}
\{\partial_s,\partial_t,s\partial_s,t\partial_t,s
\partial_t,t\partial_s,\partial_{z_1},z_1\partial_{z_1},\ldots,
\partial_{z_n-2},z_{n-2}\partial_{z_{n-2}}\}
\end{equation}

\section{Conclusions}
Superintegrability and exact solvability were defined in completely
different ways, though both have a group theoretical underpinning.
Superintegrability for a Hamiltonian system is defined by the requirement
that there be more integrals of motion than degrees of freedom
\cite{FM65}. It can be characterized by the fact that the corresponding
Schr\"odinger equation allows a nonabelian algebra of generalized
symmetries, containing an $n$-dimensional Abelian subalgebra \cite{ST01}.
Exact solvability is defined by the requirement that the energy spectrum
can be calculated algebraically \cite{TT01,Tu94,Tu88}. It can be
characterized by the fact that the Hamiltonian lies in the enveloping
algebra of a certain type of finite dimensional affine Lie algebra. It was
conjectured
\cite{TT01} that all maximally superintegrable systems are exactly
solvable. In this article we have confirmed the conjecture for the
considered integrable system in $E_n$.

The exact connection between superintegrability and exact solvability
remains an open problem.

\section*{Acknowledgements}

The research reported in this article was performed while P.W. was
visiting the Departamento de F\'{\i}sica Te\'orica II of Universidad
Complutense of Madrid. The final version was written while he was visiting
the Isaac Newton Institute in Cambridge. He thanks both Institutions for
their hospitality and support. The research of P.W. was partially
supported by research grants from NSERC of Canada and FCAR du
Qu\'ebec. The research of M.A.R was partially supported by DGES
Grant PB98-0821 of Spain.


\begin{references}

\bibitem{JH40} J. Jauch and E. Hill, On the problem of degeneracy in
quantum mechanics. Phys. Rev. {\bf 51}, 641--645 (1940).

\bibitem{MS96} M. Moshinsky and Yu. F. Smirnov, {\it The Harmonic
Oscillator in Modern Physics.\/} Harwood, Amsterdam, (1996).

\bibitem{Pa34} W. Pauli, \"Uber das Wasserstoffspektrum von Standpunk der
neuen Quanten-mechanik. Zeits. Physik {\bf 36}, 336--363 (1926).

\bibitem{Fo35} V. A. Fock, Zur Theorie des Wasserstoffatoms, Zeits. Physik
{\bf 98}, 145--154 (1935).

\bibitem{Ba36} V. Bargmann, Zur Theorie des Wasserstoffatoms, Zeits. Physik
{\bf 99}, 576--582 (1936).

\bibitem{En72} M. J. Englefield, {\it Group Theory and the Coulomb
Problem.\/} Wiley, New York (1972).

\bibitem{Mc72} H. V. MacIntosh, Symmetry and Degeneracy, 
in {\it Group Theory and its Applications},  edited by E.M. Loebl,
(Academic Press, New York, 1971), Vol. II, pp. 75-144,

\bibitem{KM76} E. G. Kalnins, W. Miller Jr, and P. Winternitz, The group
$O(4)$, separation of variables and the hydrogen atom. SIAM J. Appl.
Math. {\bf 30}, 630--664 (1976).

\bibitem{Be73} J. Bertrand, Th\'eor\`eme relatif au mouvement d'un point
attir\'e vers un centre fixe. Comptes Rendus Ac. Sci {\bf 77}, 849--853
(1873).

\bibitem{Go90} H. Goldstein, {\it Classical Mechanics\/} (Addison-Wesley,
Reading, MA, 1990).

\bibitem{FM65} I. Fri\v{s}, V. Mandrosov, J. Smorodinsky, M.
Uhl\'{\i}\v{r}, and P. Winternitz, On higher symmetries in quantum
mechanics. Phys. Lett. {\bf 16}, 354--356 (1965).

\bibitem{WS67} P. Winternitz, J. Smorodinsky, and M.
Uhl\'{\i}\v{r}, Symmetry groups in classical and quantum mechanics. Yad.
Fiz. {\bf 4}, 625 (1966), Sov. J. Nucl. Phys. {\bf 4}, 1326  (1967).

\bibitem{MS67} A. Makarov, J. Smorodinsky, Kh. Valiev, and P.
Winternitz, A systematic search for nonrelativistic systems with
dynamical symmetries. Nuovo Cimento A {\bf 52}, 1061--1084 (1967).

\bibitem{Ev90a} N. W. Evans, Superintegrability in classical mechanics.
Phys. Rev. A {\bf 41}, 5666--5676 (1990).

\bibitem{Ev91} N. W. Evans, Group theory of the Smorodinsky-Winternitz
system. J. Math. Phys. {\bf 32}, 3369--3375 (1991).

\bibitem{Ev90b} N. W. Evans, Superintegrability of the Winternitz
system. Phys. Lett. A {\bf 147}, 483--486 (1990).

\bibitem{KM96} E. G. Kalnins, W. Miller Jr, and G. S. Pogosyan,
Superintegrability and associated polynomial solutions: Euclidean space
and the sphere in two dimensions. J. Math. Phys. {\bf 37}, 6439--6467
(1996)

\bibitem{KM00a} E. G. Kalnins, W. Miller Jr, and G. S. Pogosyan,
Completeness of multiseparability in $E_{2,C}$. J. Phys. A
{\bf 33}, 4105--4120 (2000).

\bibitem{KM00b} E. G. Kalnins, W. Miller Jr, and G. S. Pogosyan,
Completeness of multiseparability on the complex 2-sphere. J. Phys.
A {\bf 33}, 6791--6806 (2000).

\bibitem{KM01} E. G. Kalnins, W. Miller Jr,
and G. S. Pogosyan, Completeness of
multiseparability in two-dimensional constant
curvature spaces. J. Phys. A {\bf 34},
4705--4720 (2001).

\bibitem{KK01} E. G. Kalnins, J. M. Kress, and P. Winternitz,
Superintegrability in a two-dimensional space of non-constant curvature.
(LANL preprint archives math-ph/0108015). J. Math. Phys. (to appear).

\bibitem{TT01} P. Tempesta, A. V. Turbiner, and P. Winternitz, Exact
solvability of superintegrable systems. J. Math. Phys. {\bf 42}, 4248--4257
(2001)

\bibitem{Tu94} A. V. Turbiner, Lie algebras and linear operators with
invariant subspaces, in {\it Lie Algebras, Cohomologies and New Findings in
Quantum Mechanics\/}, edited by N. Kamran and P.J. Olver, (AMS,
Providence, 1994), Vol. 160, pp. 263--310.

\bibitem{Tu88} A. V. Turbiner, Quasi-exactly solvable problems and $sl(2)$
algebra. Commun. Math. Phys. {\bf 118}, 467--474 (1988).

\bibitem{ST01} M. B. Sheftel, P. Tempesta, and P. Winternitz,
Superintegrable systems in quantum mechanics and classical Lie theory.
J. Math. Phys {\bf 42}, 659--673 (2001).

\bibitem{DS93} J. Daboul, P. Slodowy, and C. Daboul, The hydrogen
algebra as a centerless twisted Kac-Moody algebra. \	Phys. Lett. B {\bf
317}, 321--328 (1993).

\bibitem{DD98} C. Daboul and J. Daboul, From hydrogen atom to
generalized Dynkin diagrams. Phys. Lett. B {\bf 425} 135--144 (1998).

\bibitem {WF65} P. Winternitz and I. Fri\v{s}, Invariant expansions of
relativistic amplitudes and subgroups of the proper Lorentz group.
Sov. J. Nucl. Phys {\bf 1}, 636--643 (1965).

\bibitem{WL68} P. Winternitz, I.
Lukac, and Ya. Smorodinskii, Quantum numbers in the little groups of the
Poincar\'e group. Sov. J. Nucl. Phys {\bf 7}, 139--145 (1968).

\bibitem{PW73} J. Patera and P. Winternitz, A new basis for
representations of the rotation group. Lam\'e and Heun polynomials. J.
Math. Phys. {\bf 14}, 1130--1139 (1973).

\bibitem{MP91} W. Miller Jr, J. Patera, and P. Winternitz, Subgroups of
Lie groups and separation of variables. J. Math. Phys. {\bf 22}, 251--260
(1991).

\bibitem{Mi97} W. Miller Jr, {\it Symmetry and the Separation of
Variables\/} (Addison-Wesley, Reading, MA, 1997).

\bibitem{Ka86} E. G. Kalnins, {\it Separation of Variables for Riemannian
Spaces of Constant Curvature\/} (Longman, Burnt Mill, 1986).

\bibitem{HG84} J. Hietarinta, B. Grammaticos, B. Dorizzi, and A. Ramani,
Coupling-constant metamorphosis and duality between integrable
Hamiltonian systems. Phys. Rev. Lett. {\bf 53}, 1707--1710
(1984).

\bibitem{GK93} A. Gonz\'alez-L\'opez, N. Kamran, and P.J.
Olver, Normalizability of One-dimensional Quasi-Exactly Solvable
Schr\"odinger Operators. Commun. Math. Phys. {\bf 153}, 117--146 (1993).

\bibitem{Sh89} M. Shifman, New findings in quantum mechanics (Partial
algebraization of the spectral problem).  Int. J. Mod.
Phys. A {\bf 4}, 2897--2852 (1989).

\bibitem{Vi68} N. Ya. Vilenkin, {\it Special Functions and the Theory of
Group Representations\/} (AMS, Providence, R.I., 1968).

\bibitem{VK65} N. Ya. Vilenkin, G. I. Kuznetsov, and Ya. A. Smorodinskii,
Eigenfunctions of the Laplace operator realizing representations of
the groups $U(2)$, $SU(2)$, $SO(3)$, $U(3)$ and $SU(3)$ and the symbolic
method. Sov. J. Nucl. Phys. {\bf 2} 645--655 (1965).

\bibitem{IP99} A. A. Izmest'ev, G. S. Pogosyan, A. N. Sissakian, and P.
Winternitz, Contractions of Lie algebras and the separation of
variables. The $n$-dimensional sphere. J. Math. Phys. {\bf 40},
1549--1573 (1999).

\bibitem{PW01} G. S. Pogosyan and P. Winternitz, Separation of variables
and subgroup bases on n-dimensional hyperboloids (to be published).
\end{references}
\end{document}